\documentclass[twocolumn,aps,showpacs,amsfonts,amsmath,amssymb,superscriptaddress,floatfix]
{revtex4}
\usepackage[dvips]{graphicx}

\newcommand{\be}{\begin{equation}}
\newcommand{\ee}{\end{equation}}
\newcommand{\affA}{%
     Communications Research Laboratory,
     Koganei, Tokyo 184-8795, Japan}
\newcommand{\affB}{%
     CREST, Japan Science and Technology Corporation}
\newcommand{\affC}{%
     Advanced Research Laboratory, Hitachi Ltd.,
     Hatoyama, Saitama 350-03, Japan}

\begin{document}
\title{Unambiguous quantum state filtering}
\author{Masahiro Takeoka}
\affiliation{\affA}
\affiliation{\affB}
\author{Masashi Ban}
\affiliation{\affC}
\author{Masahide Sasaki}
\affiliation{\affA}
\affiliation{\affB}
%
\begin{abstract}

In this paper, we consider 
the generalized measurement where 
one particular quantum signal is 
unambiguously extracted from 
a set of non-commutative quantum signals and 
the other signals are filtered out. 
Simple expressions for 
the maximum detection probability and its POVM 
are derived. 
We applyl such unambiguous quantum state filtering 
to evaluation of the sensing of decoherence channels. 
The bounds of the precision limit for a given quantum state of probes and 
possible device implementations are discussed. 

\end{abstract}
\pacs{03.65.Ta, 42.50.Dv}
\date{\today}
\maketitle

\section{Introduction} 
Discrimination of non-commutative quantum states is 
one of the central issues 
in the field of quantum information processing.
Since quantum mechanics does not allow us to 
discriminate non-commutative states perfectly, 
several quantum measurement strategies 
have been studied for 
various figures of merits, such as 
average error probability \cite{Helstrom_QDET}, 
mutual information \cite{Davies78,Sasaki99}, and success probability of 
unambiguous state discrimination 
\cite{Chefles98-1,Chefles98-2,Eldar03,Herzog02,Sun02,Rudolph03}. 
These studies have been motivated not only by 
academic interest, but also more technological interests 
from the viewpoints of quantum communication, quantum cryptography, 
and interferometric sensing.

In this paper, we discuss an alternative class of quantum measurement, 
where a signal is unambiguously extracted when 
it is in a particular target state, i.e., 
unambiguous quantum state filtering.
Suppose a signal set consists of arbitrary quantum states 
$\hat{\rho}_0$ and $\hat{\rho}_1$, and 
the signals are detected by the measurement operators 
$\hat{\Pi}_0$ and $\hat{\Pi}_1$. 
We consider the measurement which maximizes 
the success probability of detecting $\hat{\rho}_0$, 
$P={\rm Tr\,}[\hat{\Pi}_0 \hat{\rho}_0]$, 
under the condition that $\hat{\rho}_1$ never be 
detected incorrectly as $\hat{\rho}_0$, i.e., 
${\rm Tr\,}[\hat{\Pi}_0 \hat{\rho}_1]=0$. 
This is the special case of Neyman-Pearson 
hypothesis testing \cite{NeymanPearson33}, namely, 
a strategy to maximize the success probability 
(to follow the conventional theory, we call it 
the `detection probability') 
while keeping the false-alarm probability equal to zero.

\begin{figure}
\begin{center}
\includegraphics[width=0.40\textwidth]{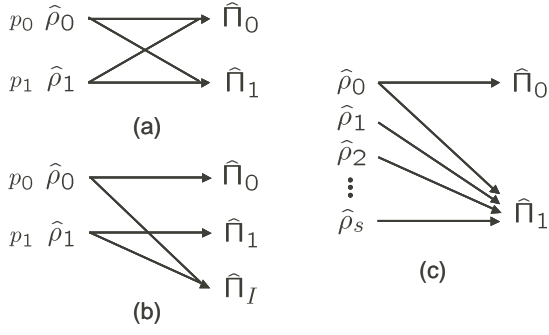}
\end{center}
\caption{\label{fig:Z-measurement}
Various generalized quantum measurement scenarios: 
(a) minimum-error state discrimination, 
(b) unambiguous state discrimination and 
(c) unambiguous state filtering. 
Here, $p_i$'s in (a) and (b) are the prior probabilities of signals.
}
\end{figure}

The Neyman-Pearson approach is effective 
for a decision problem where 
the prior probabilities of signals are unreliably known or unknown. 
Its quantum version 
has recently been applied to quantum interferometric sensing 
\cite{Paris97,D'Ariano02,Takeoka02}. 
The problem of the quantum Neyman-Pearson approach is, however, that 
the analytical solution is only known for pure state signals 
\cite{Helstrom_QDET}. 
Our formulation 
includes mixed state signals and 
we show that the maximum detection probability and 
its corresponding detection operators are given 
by quite simple expressions. 
We also generalize the problem to the case of more than two signals. 
Figure~\ref{fig:Z-measurement} compares our state filtering scenario to 
the other measurement scenarios discussed so far. 
It is stressed again that the cost of our testing scenario does not include 
the prior probabilities of the signals. 
We also note that the pure state filtering scenarios with 
Bayesian hypothesis testing have been discussed in context of 
minimum-error and unambiguous discriminations 
of quantum signal subsets \cite{Herzog02,Sun02}, 
and the latter one was recently recasted by the more general theory 
of mixed state unambiguous discrimination \cite{Rudolph03}.

After discussing the formulation of unambiguous state filtering, 
we apply it to the sensing of quantum channels consisting of 
non-unitary operations as an example of applications. 
This scenario would be important for practical applications 
because various kinds of decoherence 
in quantum channels are described by non-unitary operators. 
To sense small decoherence parameters of such channels, 
we need to detect the mixed state probes appropriately. 
We discuss the detection strategy based on unambiguous state filtering 
and propose possible experimental implementations.

The paper is organized as follows. 
In Sec.~II, the general formulation of the problem treated 
in this paper is given. 
In Sec.~III and IV, we apply our formalism 
to the problem of non-unitary quantum channel sensing. 
As concrete examples, we consider the sensing of 
two practically important operations, i.e., 
depolarizing and linear loss 
in discrete and continuous variable quantum channels, 
respectively. 
Section~V contains our concluding remarks.

\section{Optimal measurement for unambiguous state filtering}
In this section, we derive a rigorous formulation of 
the optimal measurement for our problem. 
We discuss first the filtering strategy for 
a binary signal set 
and then generalize it to 
the case of more than two signals. 
The generalization can be done straightforwardly.

Let us review the problem. 
Arbitrary quantum signals $\hat{\rho}_0$ and $\hat{\rho}_1$ are 
detected by the positive operator-valued measure (POVM) 
$\{ \hat{\Pi}_0,\,\hat{\Pi}_1 \}$, 
where $\hat{\Pi}_0+\hat{\Pi}_1=\hat{I}$, 
and our task is to maximize the detection probability, 
$P={\rm Tr\,}[\hat{\Pi}_0 \hat{\rho}_0]$, 
under the condition that 
${\rm Tr\,}[\hat{\Pi}_0 \hat{\rho}_1]=0$. 
We note that this kind of measurement 
is described by the Z-channel model and 
was first 
discussed by Kennedy 
for the detection of binary phase-shift keyed coherent signals 
with quasi-minimum average error probability \cite{Kennedy73}.

We denote the Hilbert spaces 
supporting $\hat{\rho}_0$ and $\hat{\rho}_1$ 
by ${\cal H}_0$ and ${\cal H}_1$, respectively, 
and their union by 
${\cal H} = {\cal H}_0 \cup {\cal H}_1$. 
We assume that the dimensions of ${\cal H}$ and ${\cal H}_1$ are 
$n=\dim {\cal H} > 0$ and $m=\dim {\cal H}_1 > 0$, respectively, 
where obviously $n \ge m$. 
Since we can limit the support spaces of 
$\hat{\Pi}_0$ and $\hat{\Pi}_1$ 
within the Hilbert space ${\cal H}$, 
these operators are expressed as 
\begin{eqnarray}
\label{eq:POVMSpectralDecomp.0}
\hat{\Pi}_0 & = & 
\sum_{k=1}^{n} \pi_k |\phi_k\rangle\langle\phi_k|\ , \\
\label{eq:POVMSpectralDecomp.1}
\hat{\Pi}_1 & = & 
\sum_{k=1}^{n} (1-\pi_k) |\phi_k\rangle\langle\phi_k|\ ,
\end{eqnarray}
where the set 
$\{ |\phi_1\rangle, |\phi_2\rangle,...,|\phi_n\rangle \}$
is the complete orthonormal basis set in ${\cal H}$ and 
the positivity of detection operators requires $0 \le \pi_k \le 1$ 
for all $k$.
As a consequence, our problem is now to decide 
the eigenvalues $\{ \pi_k \}$ and eigenvectors $\{ |\phi_k\rangle \}$ 
of these operators.

From Eq.~(\ref{eq:POVMSpectralDecomp.0}), the necessary condition 
for our measurement is now rewritten as 
\begin{equation}
\label{eq:FalseAlarmProb.}
{\rm Tr\,}[\hat{\Pi}_0 \hat{\rho}_1] =
\sum_{k=1}^{n} \pi_k \langle\phi_k|\hat{\rho}_1|\phi_k\rangle
=0\ .
\end{equation}
Since $\hat{\rho}_1$ is supported by the $m$-dimensional Hilbert space 
${\cal H}_1$, its spectral decomposition is given by 
\begin{equation}
\label{eq:rho1SpectralDecomp.}
\hat{\rho}_1 = \sum_{\mu=1}^{m} \lambda_{\mu} 
|\psi_{\mu}\rangle\langle\psi_{\mu}|\ ,
\end{equation}
where $\lambda_{\mu}>0$ and 
$\sum \lambda_{\mu} = 1$ ($\mu=1,2,...,m$).
The set 
$\{ |\psi_1\rangle, |\psi_2\rangle,...,|\psi_m\rangle \}$ 
is the complete orthonormal basis set in ${\cal H}_1$.
Equations (\ref{eq:FalseAlarmProb.}) and (\ref{eq:rho1SpectralDecomp.}) 
give the condition, 
\begin{equation}
\label{eq:ConditionPiPhi}
\sum_{k=1}^{n} \sum_{\mu=1}^{m} \lambda_{\mu} \pi_{k}
|\langle\phi_k|\psi_{\mu}\rangle|^2 
= \sum_{\mu=1}^{m} \lambda_{\mu} \left(
\sum_{k=1}^{n} \pi_k |\langle\phi_k|\psi_{\mu}\rangle|^2 \right) =0\ .
\end{equation}
In other words, the condition 
\begin{equation}
\label{eq:pi=0phi=0}
\pi_k = 0 \quad {\rm or} \quad \langle\phi_k|\psi_{\mu}\rangle = 0\ ,
\end{equation}
must be satisfied 
for all $k$ and $\mu$ ($k=1,2,...,n;\, \mu=1,2,...,m$).
Since the set 
$\{ |\phi_1\rangle, |\phi_2\rangle,...,|\phi_n\rangle \}$
is complete and orthonormal in ${\cal H}={\cal H}_0 \cup {\cal H}_1$, 
if $\langle\phi_k|\psi_{\mu}\rangle = 0$ for all $k$, 
then $|\psi_{\mu}\rangle=0$, 
and this contradicts the assumption $m = \dim {\cal H}_1 > 0$.
Eventually, there is at least one index $k_{\mu}$, 
such that $\pi_{k_{\mu}}=0$ for each $\mu$, that is, 
at least $m$ eigenvalues take a zero value,
\begin{equation}
\label{eq:pi0} 
\pi_{k_1}=\pi_{k_2}= ... =\pi_{k_m}=0\ .
\end{equation}

Now let us maximize the detection probability $P$,
\begin{equation}
\label{eq:P}
P = {\rm Tr\,}[\hat{\Pi}_0 \hat{\rho}_0] 
= \sum_{k=1}^{n} \pi_k 
\langle \phi_k | \hat{\rho}_0 | \phi_k \rangle\ .
\end{equation}
Since $\langle\phi_k|\hat{\rho}_0|\phi_k\rangle \ge 0$, 
we should choose $\pi_k$ ($k=1,2,...,n$) to be as large as possible 
to maximize $P$.
From Eq.~(\ref{eq:pi0}) and 
$0 \le \pi_k \le 1$, we find that $P$ is maximized when
\begin{equation}
\label{eq:condition}
\left\{ 
\begin{array}{l}
\pi_{k_1}=\pi_{k_2}=...=\pi_{k_m}=0, \\
\pi_k = 1 \quad (k \ne k_1,k_2,...,k_m)\ ,
\end{array} \right. 
\end{equation}
is satisfied.

Next, we derive the relation between 
$\{ |\phi_k\rangle \}$ and $\{ |\psi_{\mu}\rangle \}$. 
Let us consider the case that every index $k_{\mu}$ is 
not equal to each other, i.e., 
$\mu_i \ne \mu_j \to k_{\mu_i} \ne k_{\mu_j}$ \cite{comment1}.
Since Eq.~(\ref{eq:condition}) means that, 
for each $\mu$, there is only one index $k_{\mu}$ 
such that $\langle\phi_{k_{\mu}}|\psi_{\mu}\rangle \ne 0$, 
$|\psi_{\mu}\rangle$ can be expressed as 
\begin{equation}
\label{eq:Psi=phi}
|\psi_{\mu}\rangle = \sum_{k=1}^{n} g_{\mu} (k) |\phi_k\rangle
= |\phi_{k_{\mu}}\rangle \ .
\end{equation}
As a consequence, the detection operator $\hat{\Pi}_0$ 
\begin{eqnarray}
\label{eq:Pi_0}
\hat{\Pi}_0 & = & \sum_{k=1}^{n} 
|\phi_k\rangle\langle\phi_k|
\quad (k \ne k_1,k_2,...,k_m)
\nonumber\\
& = & 
\hat{I} - \sum_{\mu=1}^{m} 
|\psi_{\mu}\rangle\langle\psi_{\mu}|\ .
\end{eqnarray}
Then we arrive at the optimal POVM, 
\begin{equation}
\label{eq:Pi_0Pi_1}
\hat{\Pi}_0 = \hat{I} - 
\sum_{\mu=1}^{m} |\psi_{\mu}\rangle\langle\psi_{\mu}|,
\quad
\hat{\Pi}_1 = 
\sum_{\mu=1}^{m} |\psi_{\mu}\rangle\langle\psi_{\mu}|\ .
\end{equation}
Apparently, if $n=m$, then $\hat{\Pi}_0=0$, that is, 
when the support of $\hat{\rho}_0$ is included 
in that of $\hat{\rho}_1$ (${\cal H}_0 \subseteq {\cal H}_1$), 
unambiguous detection of $\hat{\rho}_0$ is impossible \cite{comment2}.
On the other hand, $P=1$ can be obtained when 
${\cal H}_0 \cap {\cal H}_1 = \emptyset$ or, equivalently, 
$\hat{\rho}_0 \hat{\rho}_1 = \hat{\rho}_1 \hat{\rho}_0 =0$.
When we are given a pure state 
$\hat{\rho}_1=|\psi\rangle\langle\psi|$,
the optimal POVM is given by 
\begin{equation}
\label{eq:POVM_pure_rho_1}
\hat{\Pi}_0 = \hat{I} - |\psi\rangle\langle\psi| \quad
\hat{\Pi}_1 = |\psi\rangle\langle\psi| \ .
\end{equation}
This POVM is the same as 
that for pure state discrimination 
in \cite{Takeoka02}. 
This means that the physical implementation scheme 
proposed in \cite{Takeoka02} 
can be applied to mixed state signals.
We further discuss this issue in the later section.

Finally, let us consider the following more general problem: 
we are given $s+1$ quantum states, 
$\hat{\rho}_0,\hat{\rho}_1,...,\hat{\rho}_s$, and 
to find the measurement such that 
(1) $\hat{\rho}_1,\hat{\rho}_2,...,\hat{\rho}_s$ never be 
recognized as $\hat{\rho}_0$, and 
(2) the detection probability of $\hat{\rho}_0$ be maximized 
(Fig.~\ref{fig:Z-measurement}(b)). 
For this purpose, we rewrite Eq.~(\ref{eq:Pi_0Pi_1}) as 
\begin{equation}
\label{eq:Pi_0Pi_1re}
\hat{\Pi}_0 = \hat{I}_{{\cal H}_0 \cup {\cal H}_1} - 
\hat{I}_{{\cal H}_1},
\quad
\hat{\Pi}_1 = \hat{I}_{{\cal H}_1}\ ,
\end{equation}
where $\hat{I}_{{\cal H}}$ is the identity operator defined 
in the Hilbert space ${\cal H}$. 
When we denote the support space of $\hat{\rho}_k$ as 
${\cal H}_k$, 
Eq.~(\ref{eq:Pi_0Pi_1re}) is easily generalized as 
\begin{equation}
\label{eq:s+1POVM}
\begin{array}{ccl}
\hat{\Pi}_0 & = & \hat{I}_{
{\cal H}_0 \cup {\cal H}_1 \cup {\cal H}_2 \cup \cdot\cdot\cdot 
\cup {\cal H}_s
} - 
\hat{I}_{
{\cal H}_1 \cup {\cal H}_2 \cup \cdot\cdot\cdot 
\cup {\cal H}_s }\ ,
\\
\hat{\Pi}_1 & = & \hat{I}_{
{\cal H}_1 \cup {\cal H}_2 \cup \cdot\cdot\cdot 
\cup {\cal H}_s}\ .
\end{array}
\end{equation}
The maximum detection probability $P$ is then given by 
\begin{equation}
\label{eq:P_s+1}
P = {\rm Tr\,}[\hat{\Pi}_0 \hat{\rho}_0]
= 1 - \sum_{k=1}^M 
\langle \Psi_k |\hat{\rho}_0| \Psi_k \rangle,
\end{equation}
where $\{ |\Psi_1\rangle, |\Psi_2\rangle, ..., |\Psi_M\rangle \}$ 
is the complete orthonormal vector set of 
the Hilbert space 
${\cal H}_1 \cup {\cal H}_2 \cup \cdot\cdot\cdot \cup {\cal H}_s$.

\section{Application I: Sensing of depolarization
in a discrete channel}
In the following two sections, 
we apply our measurement strategy 
to sensing applications.
One of the simplest quantum sensing scenarios can be described 
as follows \cite{Paris97}: 
A probe field, initially prepared in $\hat{\rho}_1$, 
travels through a sample in which 
the probe field may or may not be perturbed. 
A perturbation is generally described by 
a completely positive (CP) map ${\cal L}$. 
If a perturbation occurs, the probe field is 
then modified to $\hat{\rho}_0 = {\cal L} \, \hat{\rho}_1$. 
Eventually, we may have an unperturbed state $\hat{\rho}_1$ or 
a perturbed state $\hat{\rho}_0$ as output. 
The task is to maximize the probability that 
the signal will be unambiguously detected when it suffers 
a perturbation ${\cal L}$.

First, let us consider a discrete quantum channel consisting of 
an $n$-dimensional depolarizing channel 
${\cal L}_D$ which is characterized by 
the depolarizing probability $p$ \cite{NielsenChuang}.
This example is simple and instructive to see 
how non-classical probes and appropriate generalized measurements 
enhance the detection probability.
We probe the channel, in which depolarization  
${\cal L}_D$ may or may not occur,
using a pure state input 
$\hat{\rho}_1 = |\psi\rangle\langle\psi|$.
If depolarizing occurs, the output state is given by 
\cite{NielsenChuang}
\begin{equation}
\label{eq:CPmap_depolarizing}
\hat{\rho}_0 = {\cal L}_D |\psi\rangle\langle\psi|
= (1-p) |\psi\rangle\langle\psi| + \frac{p}{n} \hat{I} \ .
\end{equation}
Then, following Eqs.~(\ref{eq:P}) and (\ref{eq:POVM_pure_rho_1}), 
the maximum detection probability of inferring the state $\hat{\rho}_0$ 
is given by 
\begin{equation}
\label{eq:CPmap_P}
P_D = 1 - \langle\psi| 
\left( {\cal L}_D |\psi\rangle\langle\psi| \right) 
|\psi\rangle
= \frac{n-1}{n} p \ ,
\end{equation}
where $P_D$ is independent of the input quantum state.

Next, let us consider an entangled input. 
An arbitrary entangled state in an $n$-dimensional space is 
described by 
$|\Psi\rangle = \sum_k \sqrt{\lambda_k} 
|k\rangle \otimes |k\rangle$ $(k=1,2,...,n$, $\sum_k \lambda_k = 1)$ 
and now one part of the state is incident on the channel. 
The maximum detection probability of inferring 
the state 
$\hat{\rho}_0 = ({\cal L}_D \otimes {\cal I}) |\Psi\rangle\langle\Psi|$ 
is then given by 
\begin{equation}
\label{eq:CPmap_Pentangle}
P_D^{\rm ent} = 1 -
\langle\Psi| [ 
({\cal L}_D \otimes {\cal I}) |\Psi\rangle\langle\Psi| ]
|\Psi\rangle 
= \left( 1-\frac{1}{n} \sum_{k=1}^{n} \lambda_k^2 \right) p \ .
\end{equation}
Here 
the maximum value of $P_D^{\rm ent}$ is obtained when 
$\lambda_1 = \lambda_2 = \cdot\cdot\cdot = \lambda_n = 1/n$, 
that is, the state is maximally entangled. 
On the other hand, when one of $\lambda_k$ is 1 and the others are 0 
(i.e., the state is not entangled), 
$P_D^{\rm ent}$ takes its minimum value and becomes equal to 
$P_D$ in Eq.~(\ref{eq:CPmap_P}). 
Therefore, we see that $P_D^{\rm ent} \ge P_D$ is always held, 
that is, entanglement always improves 
the detection probability. 
Use of an entangled input increases 
the Hilbert space of the output state, 
enhancing the overlap between $\hat{\rho}_0$ and $\hat{\rho}_1$, 
as has been pointed out in \cite{D'Ariano01}.

\section{Application II: Sensing of linear loss in 
a continuous variable channel} 
In this section, we examine the sensing of linear loss 
through probing by continuous variable quantum states. 
Linear loss caused by coupling between the system and 
a vacuum environment is one of common problems of decoherence 
in quantum optics and quantum information processing. 
When we assume that a probe is in a pure state 
$\hat{\rho}_1 = |\psi\rangle\langle\psi|$, 
such as a coherent state, squeezed state or 
two-mode squeezed state, 
the optimal POVM 
is given by Eq.~(\ref{eq:POVM_pure_rho_1}). 
As mentioned in Section II, 
this is the same as that described in \cite{Takeoka02}. 
Therefore, the physical implementation scheme 
for unitary operation sensing proposed in \cite{Takeoka02} 
can be directly applied to this problem. 
The purpose of this section is to clarify 
the ultimate sensing limits of linear loss 
for concretely given probe states.

The CP map of linear loss is given by \cite{QuantumOptics} 
\begin{equation}
\label{eq:CPmap_LinearLoss}
{\cal L}_{L} = 
\exp \left[ g \left( \hat{\cal K}_- - \hat{\cal K}_0 \right) \right]
\ ,
\end{equation}
where $g$ is a positive parameter and 
the superoperators $\hat{\cal K}_-$ and $\hat{\cal K}_0$ are defined by 
\begin{equation}
\label{eq:Loss_SuperOperators}
\hat{\cal K}_- \hat{X} = \hat{a} \hat{X} \hat{a}^{\dagger} \qquad
\hat{\cal K}_0 \hat{X} = {\textstyle\frac{1}{2}} 
(\hat{a} \, \hat{a}^{\dagger} \hat{X} + \hat{X} \hat{a}^{\dagger} \hat{a})
\ ,
\end{equation}
for an arbitrary operator $\hat{X}$. 
Here, $\hat{a}$ and $\hat{a}^{\dagger}$ are, respectively, 
annihilation and creation operators. 
The CP map ${\cal L}_{L}$ transforms a coherent state 
into another coherent state 
with reduced complex amplitude as 
\begin{equation}
\label{eq:CoherentStateMapping}
{\cal L}_L | \alpha \rangle \langle \beta | 
    =  E(\alpha, \beta)\,
    | \alpha \sqrt{T} \rangle \langle \beta \sqrt{T} | \ ,
\end{equation}
where $T=\exp(-g)$ corresponds to 
the transmittance of the CP map and 
\begin{equation}
\label{eq:E(alpha,beta)}
E(\alpha, \beta) =
\exp \left[ - \frac{1}{2} (1-T) 
    \left( | \alpha |^2 + | \beta |^2 
           - 2 \alpha \beta^* \right) 
  \right] \ .
\end{equation}
In this section, 
we use $R=1-T$ as a parameter of the degree of loss.

Following the previous works \cite{Paris97,D'Ariano02,Takeoka02}, 
we compare 
the ultimate sensitivities for various probes 
in terms of $R_{M}$, 
which is defined by 
the solution of Eq.~(\ref{eq:P}) for $R$. 
Here, $R_{M}$ is the minimum detectable loss 
for a fixed detection probability $P$. 
This probability may be called the acceptance probability $P_{\rm ac}$ 
and the value of $P_{\rm ac}$ should be determined 
by referring to the actual experimental conditions.

\subsection{Coherent state} 
A simple example is that the probe field is in a coherent state 
$|\psi\rangle=|\alpha\rangle$. 
The detection probability is given by 
\begin{eqnarray}
\label{eq:P_coherent}
P_L^{\rm coh} 
& = & 1 - \langle\alpha| \left(
      {\cal L}_L(R) |\alpha\rangle\langle\alpha| \right)  
      |\alpha\rangle \ , \nonumber\\ 
& = & 1 - \exp\left[ 
      -\left|\left(1-\sqrt{1-R}\right)\alpha\right|^2 \right] \ .
\end{eqnarray}
Then we find $R_M$ as 
\begin{equation}
\label{eq:R_coh}
R_{M}^{\rm coh} = 
\frac{1}{\sqrt{\langle n \rangle}} 
\sqrt{\log \frac{1}{1-P_{\rm ac}}}
\left( 2 - \frac{1}{\sqrt{\langle n \rangle}} 
\sqrt{\log \frac{1}{1-P_{\rm ac}}} \right) \ ,
\end{equation}
where $\langle n \rangle = |\alpha|^2$ is 
the average photon number of the probe field. 
We can easily find that $R_{M}^{\rm coh}$ is proportional to 
$1/\sqrt{\langle n \rangle}$ within the limit of large $\langle n \rangle$, 
which is the same as in the case of sensing unitary operations 
described in \cite{Paris97,Takeoka02}. 
On the other hand, 
when $\langle n \rangle$ is small, 
$\langle n \rangle$ has to satisfy 
\begin{equation}
\label{eq:n_min_coh}
\langle n \rangle \ge \langle n \rangle_{\rm min} = 
\log \frac{1}{1-P_{\rm ac}} \ ,
\end{equation}
since $R_{M} \le 1$. 
This is different from that for the sensing of unitary operations. 
Here, $\langle n \rangle_{\rm min}$ is 
the minimum required power for sensing. 
In other words, $\langle n \rangle_{\rm min}$ is 
the average power required to unambiguously 
detect the signal with detection probability $P_{\rm ac}$, 
when $|\psi\rangle$ is mapped to a vacuum state. 
\\

\subsection{Squeezed state} 
\begin{figure}
\begin{center}
\includegraphics[width=0.40\textwidth]{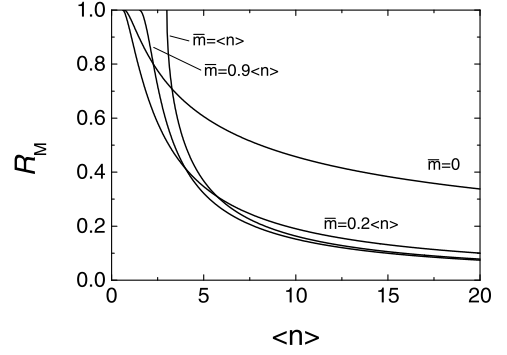}
\end{center}
\caption{\label{fig:graph1}
Minimum detectable loss $R_M$ as a function of 
total photon number $\langle n \rangle$.
$\bar{m}=\sinh^2 r$ is a function of the squeezing parameter $r$.
$P_{\rm ac}=1/2$.}
\end{figure}

As a second example let us consider the squeezed state 
$|\psi\rangle=D(\alpha)S(\zeta)|0\rangle$ 
as a probe field 
where $\zeta=r e^{i\theta}$ is the complex squeezing parameter. 
After tedious calculation, we get the detection probability 
\begin{widetext}
\begin{eqnarray}
\label{eq:P_sqzd}
P_L^{\rm sq} & = & 1 - \langle \psi| 
\left( {\cal L}_L |\psi\rangle\langle\psi| \right)
| \psi \rangle \nonumber\\
& = & 1 - 
\frac{1}{\sqrt{1 + |\nu|^2 R(2-R)}}
\exp \left[ - \frac{(1-\sqrt{1-R})^2}{1+|\nu|^2 R(2-R)}
\left\{ |\alpha|^2 + (2-R) 
\left( |\nu|^2 |\alpha|^2 + 
\frac{1}{2} \mu ( \nu^* \alpha^2 + \nu \alpha^{*2} ) \right) 
\right\} \right] \ , \nonumber\\ 
\end{eqnarray}
\end{widetext}
where 
\begin{equation}
\label{eq:mu_nu}
\mu = \cosh r , \quad \nu = e^{i\theta} \sinh r \ .
\end{equation}
Obviously, Eq.~(\ref{eq:P_sqzd}) depends on 
the phases of the coherent amplitude and squeezing. 
To maximize $P_L^{\rm sq}$, 
these factors should be optimized. 
Labeling the phase of the coherent amplitude as 
$\alpha = |\alpha|e^{i\varphi}$, 
we easily find that Eq.~(\ref{eq:P_sqzd}) is maximized 
when $2\varphi-\theta=0$ is held. 
Now we can take $\varphi=\theta=0$ (amplitude squeezing) 
without loss of generality. 
Then Eq.~(\ref{eq:P_sqzd}) is simplified to 
\begin{widetext}
\begin{equation}
\label{eq:P_squeezed}
P_L^{\rm sq} = 1 - 
\frac{1}{\sqrt{1+\sinh^2 r\,R(2-R)}}
\exp \left[ - \frac{(1-\sqrt{1-R})^2}{1+\sinh^2 r\,R(2-R)}
(1-e^{-r} \sinh r \,R)e^{2r} \alpha^2 \right] \ .
\end{equation}
\end{widetext}
In the following, we apply the power constraint condition, 
$\langle n \rangle = \bar{n}+\bar{m}$, 
where $\langle n \rangle$ is the total average number of photons, 
$\bar{n}=|\alpha|^2$ and $\bar{m}= \sinh^2 r$. 
In Fig.~\ref{fig:graph1}, 
$R_M^{\rm sq}$ with $P_{\rm ac}=1/2$ 
for a given $\langle n \rangle$ 
is plotted with $\bar{m}=0$ (coherent state), 
$\bar{m}=0.2\langle n \rangle$, $\bar{m}=0.9\langle n \rangle$, 
and $\bar{m}=\langle n \rangle$ (squeezed vacuum). 
This figure shows that 
the optimal power distribution to $\bar{n}$ and $\bar{m}$ 
depends on the total power $\langle n \rangle$ of the probe field. 

Before further considering the optimization of 
power distribution for a squeezed input, 
we give analytical expressions for some limited cases. 
When the power for the probe field is fully used for 
squeezing ($\bar{m}=\langle n \rangle$), 
i.e., the probe is in a squeezed vacuum state, 
$P_L$ is simply given by 
\begin{equation}
\label{eq:P_SV}
P_L^{\rm sv} = 1 - 
\frac{1}{ \sqrt{1+ \langle n \rangle R(2-R)} } \ .
\end{equation}
Then we find $R_M$ and the minimum required power as 
\begin{equation}
\label{eq:R_M_SV}
R_M^{\rm sv} = 1 - 
\sqrt{1-\frac{1}{\langle n \rangle} \left\{
\left( \frac{1}{1-P_{\rm ac}} \right)^2 - 1 \right\} } \ ,
\end{equation}
and 
\begin{equation}
\label{eq:n_SV}
\langle n \rangle_{\rm min} 
= \left( \frac{1}{1-P_{\rm ac}} \right)^2 - 1 \ ,
\end{equation}
respectively. 
Within the limit of large $\langle n \rangle$, 
$R_M^{\rm sv}$ is approximately proportional to $\langle n \rangle$ as 
\begin{equation}
\label{eq:R_M_SVapprox}
R_M^{\rm sv} \approx \frac{1}{2\langle n \rangle}
\left\{ \left( \frac{1}{1-P_{\rm ac}} \right)^2 -1 \right\} \ .
\end{equation}

On the other hand, when we assume $\bar{n} \gg \bar{m}$, 
which is important from a practical point of view, 
Eq.~(\ref{eq:P_squeezed}) can be approximated to 
\begin{equation}
\label{eq:P_BS}
P_L^{\rm sq} \approx 1 - 
\exp \left[ - (1-\sqrt{1-R})^2 e^{2r} \alpha^2 \right] \ ,
\end{equation}
and then 
\begin{equation}
\label{eq:R_M_BS}
R_M^{\rm sq} \approx \frac{1}{e^r \sqrt{\bar{n}}} 
\sqrt{\log \frac{1}{1-P_{\rm ac}}} \left( 2 - 
\frac{1}{e^r \sqrt{\bar{n}}} 
\sqrt{\log \frac{1}{1-P_{\rm ac}}} \right) \ ,
\end{equation}
which shows that a bright squeezed probe field 
improves the minimum detectable loss by a factor of $e^r$.

\begin{figure}
\begin{center}
\includegraphics[width=0.40\textwidth]{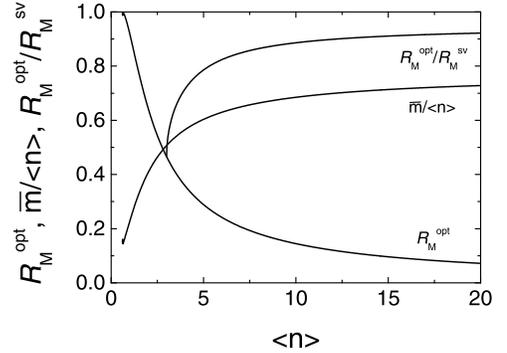}
\end{center}
\caption{\label{fig:graph2}
Minimum detectable loss $R_M$ for the squeezed probe in which 
the power distribution to $\bar{n}$ and $\bar{m}$ are optimized.
For reference, the corresponding power distribution 
$\bar{m}/\langle n \rangle$ and the ratio of $R_M$ between 
the optimized squeezed probe and the squeezed vacuum probe 
are plotted. $P_{\rm ac}=1/2$.
}
\end{figure}

Figure~\ref{fig:graph2} shows 
$R_M^{\rm opt}$ with $P_{\rm ac}=1/2$, 
where the power distribution 
for squeezing is numerically optimized. 
For reference, the optimal power ratio 
$\bar{m}/\langle n \rangle$ and 
the ratio between 
$R_M^{\rm opt}$ and $R_M^{\rm sv}$ are plotted 
in the same figure. 
We found that $R_M^{\rm opt}$ asymptotically 
reached about 92\% of $R_M^{\rm sv}$, 
which means that $R_M^{\rm opt}$ is also proportional to 
$\langle n \rangle$ 
within the limit of large $\langle n \rangle$. 
The minimum required power is about 
$\langle n \rangle_{\rm min} = 0.60$, 
which is slightly better than that of coherent state probing 
($\langle n \rangle_{\rm min} = 0.69$).


\begin{figure}
\begin{center}
\includegraphics[width=0.40\textwidth]{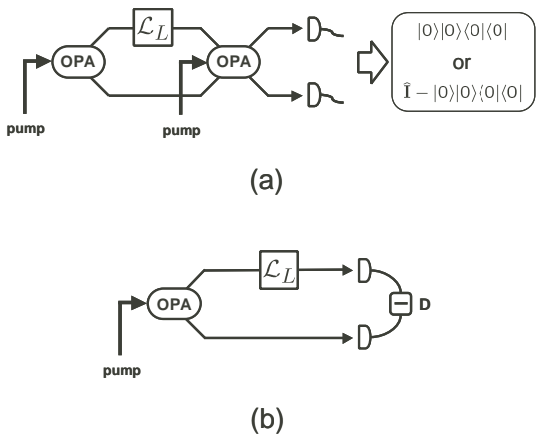}
\end{center}
\caption{\label{fig:TMSVscheme}
Sensing of linear loss by a two-mode squeezed vacuum probe and
(a) optimal measurement or 
(b) measurement of the photon number difference.
}
\end{figure}

\subsection{Two-mode squeezed vacuum}
Our final example is a probe field consisting of 
the two-mode squeezed vacuum state, 
\begin{equation}
\label{eq:TMSV}
|\Psi\rangle = \sqrt{1-\lambda^2} 
\sum_{n=0}^{\infty} \lambda^n |n\rangle |n\rangle \ ,
\end{equation}
where $\lambda=\tanh r$ and we assume that 
the squeezing parameter $r$ is real and positive 
without loss of generality. 
Let us consider the sensing scheme in which 
one part of the two-mode squeezed vacuum state 
is incident on a tested channel and then 
all of the output modes are measured collectively. 
The optimal POVM is described by $\{ \hat{\Pi}_0 = 
\hat{I} - |\Psi\rangle\langle\Psi|,\, 
\hat{\Pi}_1 = |\Psi\rangle\langle\Psi| \}$ and thus 
we obtain the detection probability as 
\begin{eqnarray}
\label{eq:P_tmsv}
P_L^{\rm tmsv} & = & 1 - \langle\Psi| \left[
\left( {\cal L}_L \otimes {\cal I} \right) 
|\Psi\rangle\langle\Psi| \right] |\Psi\rangle
\nonumber\\ 
& = & 1 - \left(
\frac{1-\lambda^2}{1-\lambda^2\sqrt{1-R}} \right)^2 \ .
\end{eqnarray}
Note that this measurement can be physically implemented through  
the same strategy as used by \cite{Takeoka02}, for example, 
through a system consisting of 
parametric three-wave mixing and photo-detection 
to discriminate between zero and non-zero photons, 
as shown in Fig.~\ref{fig:TMSVscheme}(a). 
From Eq.~(\ref{eq:P_tmsv}), we find $R_M$ as 
\begin{eqnarray}
\label{eq:R_tmsv}
R_M^{\rm tmsv} & = & \frac{1}{\langle n \rangle}
\left( \frac{1}{\sqrt{1-P_{\rm ac}}} - 1 \right)
\nonumber\\ & & \times \,
\left\{ 2 - \frac{1}{\langle n \rangle}
\left( \frac{1}{\sqrt{1-P_{\rm ac}}} - 1 \right) \right\} \ ,
\end{eqnarray}
where the minimum required power is 
\begin{equation}
\label{eq:n_tmsv}
\langle n \rangle_{\rm min} = \frac{1}{\sqrt{1-P_{\rm ac}}} - 1 \ .
\end{equation}
Here, $\langle n \rangle=\lambda^2/(1-\lambda^2)$ is 
the average power incident on a tested channel. 
Obviously, within the limit of large $\langle n \rangle$, 
$R_M^{\rm tmsv}$ is almost proportional to $1/\langle n \rangle$.

Finally, we briefly discuss another strategy that is 
based on measurement of the photon number (photocurrent) 
difference, which works effectively 
in some interferometric schemes using entanglement \cite{D'Ariano02}.
The schematic is shown in Fig.~\ref{fig:TMSVscheme}(b). 
Here the state is discriminated by observing 
whether the difference in the photon number between two photo-detectors 
is zero or non-zero, i.e., 
by the set of measurement operators 
$\{ \hat{\Pi}_0 = \hat{I} - 
\sum_{n} |n\rangle|n\rangle\langle n|\langle n| ,\, 
\hat{\Pi}_1 = \sum_{n} |n\rangle|n\rangle\langle n|\langle n| \}$. 
Although this is not optimal since the support space of 
$\sum_{n} |n\rangle|n\rangle\langle n|\langle n|$ is obviously 
larger than that of Eq.~(\ref{eq:TMSV}), 
to find the precision limit of this simpler measurement scheme 
is still interesting from a practical point of view. 
The detection probability is calculated to be 
\begin{eqnarray}
\label{eq:P_photoncounting}
P_L^{\rm tmsv} & = & 1 - \sum_{n=0}^{\infty} 
\langle n| \langle n| \left[
\left( {\cal L}_L \otimes {\cal I} \right) 
|\Psi\rangle\langle\Psi| \right] |n\rangle |n\rangle 
\nonumber\\ 
& = &
1 - \frac{1-\lambda^2}{1-\lambda^2 (1-R)} \ ,
\end{eqnarray}
and we find the minimum detectable loss and 
the minimum required power as 
\begin{equation}
\label{eq:R_photoncounting}
R_M^{\rm tmsv} = \frac{1}{\langle n \rangle}
\left( \frac{P_{\rm ac}}{1-P_{\rm ac}} \right) \ ,
\end{equation}
and 
\begin{equation}
\label{eq:n_photoncounting}
\langle n \rangle_{\rm min} 
= \frac{P_{\rm ac}}{1-P_{\rm ac}} \ ,
\end{equation}
respectively. 

\begin{figure}
\begin{center}
\includegraphics[width=0.40\textwidth]{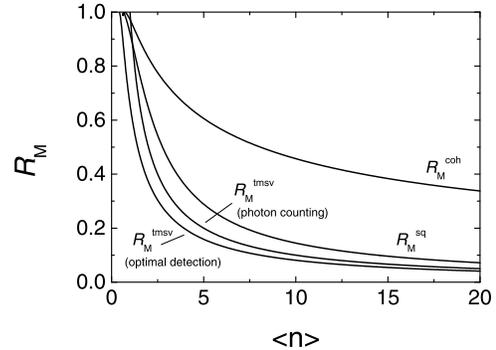}
\end{center}
\caption{\label{fig:graph3}
Minimum detectable loss $R_M$ for various probe fields. 
}
\end{figure}

To summarize, we plotted the $R_M$'s for coherent, squeezed, and 
two-mode squeezed probes 
with $P_{\rm ac}=1/2$ in Fig.~\ref{fig:graph3}. 
Again, a two-mode squeezed (entangled) probe showed better performance 
than single-mode squeezed probes. 
Since the total power for a probe is defined by 
{\it the power incident on a tested channel only}, 
the expansion of the output Hilbert space due to an entangled probe 
clearly enables greater sensitivity. 
Fortunately, we can also see that 
when the input power is sufficiently larger than 
$\langle n \rangle_{\rm min}$, 
the strategy of detecting the photon number difference 
is near optimal. 
This strategy may be simpler than the optimal one 
in regions of large $\langle n \rangle$ where 
currently available photodetectors can operate. 
However, an entangled probe provides the most drastic gain 
in regions where the input power is extremely limited, 
when it is measured by the optimal measurement scheme.

\section{Concluding remarks} 
In this paper, we have discussed the measurment scenario of 
unambiguous quantum state filtering 
where one particular signal is 
unambiguously extracted 
from a set of non-orthogonal signals. 
The maximum detection probability and 
its corresponding measurement operators are given by 
simple expressions 
although our formulation includes mixed state signals.

As an application, we have considered the sensing of decoherence 
in quantum channels. 
We applied this formalism 
to the sensing of two kinds of decoherence channels, 
a depolarizing channel and a channel which included linear loss. 
A probe field is initially in a pure state and 
the output is probably in a mixed state. 
The latter channel is especially important in practice 
since we often encounter such decoherence 
in optical quantum communication networks. 
Our results suggest that the asymptotic behavior of the precision limit 
is the same as that of unitary operation sensing, 
for example, phase shift sensing \cite{Paris97,D'Ariano02,Takeoka02}. 
On the other hand, in the region of a weak probe field, 
for probes to have non-zero detection probabilities 
there are minimum required powers. 
These results show how a non-classical, especially an entangled, probe 
and an appropriate detection scheme improve 
sensing performance.


\begin{references}


\bibitem{Helstrom_QDET}
   C.~W.~Helstrom, {\it Quantum Detection and Estimation Theory}
   (Academic Press, New York, 1976).


\bibitem{Davies78}
   E.~B.~Davies, 
   IEEE Trans.\ Inf.\ Theory\,\textbf{IT-24}, 596 (1978).


\bibitem{Sasaki99}
   M.~Sasaki, S.~M.~Barnett, R.~Jozsa, M.~Osaki, and O.~Hirota,
   Phys.\ Rev.\ A\,\textbf{59}, 3325 (1999). 


\bibitem{Chefles98-1}
   A.~Chefles, 
   Phys.\ Lett.\ A\,\textbf{239}, 339 (1998).


\bibitem{Chefles98-2}
   A.~Chefles and S.~M.~Barnett, 
   Phys.\ Lett.\ A\,\textbf{250}, 223 (1998).


\bibitem{Eldar03}
   Y.~C.~Eldar,
   IEEE Trans.\ Inf.\ Theory\,\textbf{IT-49}, 446 (2003).


\bibitem{Herzog02}
   U.~Herzog and J.~A.~Bergou, 
   Phys.\ Rev.\ A\,\textbf{65}, 050305(R) (2002). 


\bibitem{Sun02}
   Y.~Sun, J.~A.~Bergou, and M.~Hillery, 
   Phys.\ Rev.\ A\,\textbf{66}, 032315 (2002). 


\bibitem{Rudolph03}
  T.~Rudolph, R.~W.~Spekkens, and P.~S.~Turner, 
  LANL arXive quant-ph/0303071 (2003).


\bibitem{NeymanPearson33}
   J.~Neyman and E.~Pearson, 
   Philos.\ Trans.\ R.\ Soc.\ London,\ Ser.\ A\,\textbf{231}, 289 (1933).


\bibitem{Paris97}	
   M.~G.~A.~Paris,
   Phys.\ Lett.\ A\,\textbf{225}, 23 (1997). 


\bibitem{D'Ariano02}
   G.~M.~D'Ariano, M.~G.~A.~Paris, and P.~Perinotti,
   Phys.\ Rev.\ A\,\textbf{65}, 062106 (2002). 


\bibitem{Takeoka02}
  M.~Takeoka, M.~Ban, and M.~Sasaki, 
  LANL arXive quant-ph/0212037 (2002).


\bibitem{Kennedy73}
   R.~S.~Kennedy, 
   Research Laboratory of Electronics, MIT, 
   Quarterly Progress Report No.~108, 219 (1973).


\bibitem{comment1}
Suppose $k_{\mu_1}=k_{\mu_2}\,(\equiv k_0)$.
It induces 
$\langle\phi_{k_0}|\psi_{\mu_1}\rangle 
= \langle\phi_{k_0}|\psi_{\mu_2}\rangle \ne 0$ and 
$\langle\phi_{k}|\psi_{\mu_1}\rangle 
= \langle\phi_{k}|\psi_{\mu_2}\rangle = 0$ ($k \ne k_0$), 
eventually, 
$|\psi_{\mu_1}\rangle=|\psi_{\mu_2}\rangle=|\phi_{k_0}\rangle$.
Since $\{ |\psi_1\rangle,|\psi_2\rangle,...,|\psi_m\rangle \}$ is 
the set of eigenvectors of $\hat{\rho}_1$, 
the above statement is in contradiction to 
the assumption that $\hat{\rho_1}$ 
is supported by the $m$-dimensional Hilbert space ${\cal H}_1$.


\bibitem{comment2}
For example, 
although it is possible to construct the filtering measurement 
which extracts the thermal state unambiguously 
from a signal set consisting of a thermal state and 
a vacuum state, 
its converse (extraction of the vacuum state) is impossible.


\bibitem{NielsenChuang}
   M.~A.~Nielsen and I.~L.~Chuang, 
   {\it Quantum Computation and Quantum Information}
   (Cambridge University Press, Cambridge, 2000).


\bibitem{D'Ariano01}
   G.~M.~D'Ariano, P.~LoPresti, and M.~G.~A.~Paris,
   Phys.\ Rev.\ Lett.\,\textbf{87}, 270404 (2001). 


\bibitem{QuantumOptics}
   D.~F.~Walls and G.~J.~Milburn,
   {\it Quantum Optics}
   (Springer-Verlag, Berlin, 1994).






\end{references}
\end{document}